\documentclass[]{spie}

\usepackage[]{graphicx}
\usepackage[numbers]{natbib}
\usepackage{amsmath,amsfonts,amssymb}
\usepackage[outercaption]{sidecap}
\usepackage[dvipsnames]{xcolor}
\usepackage{multirow}
\usepackage{tcolorbox}
\usepackage{colortbl}
 \usepackage{xcolor}

\def\arcsec{\hbox{$^{\prime\prime}$}}
\def\deg{\hbox{$^\circ$}}

\def\lae{\mathrel{\raise .4ex\hbox{\rlap{$<$}\lower 1.2ex\hbox{$\sim$}}}}
\def\gae{\mathrel{\raise .4ex\hbox{\rlap{$>$}\lower 1.2ex\hbox{$\sim$}}}}

\definecolor{myred}{cmyk}{0., .99, 0.78, 0.45, 1.00}

\newcommand{\axaf}{\mbox{\em Chandra\/}}

\newcommand{\rs}{{\em REDSoX}}

\newcommand{\sxp}{{\em GOSoX}}

\newcommand{\ixpe}{\mbox{\em IXPE\/}}

\title{A new NASA Pioneer: the Globe Orbiting Soft X-ray Polarimeter (GOSoX)}

\author[a]{Herman L.\ Marshall}
\author[a]{Sarah N.~T.~Heine}
\author[a]{Alan Garner}
\author[a]{Sean Gunderson}
\author[a]{Hans M.\ G\"unther}
\author[b]{Ralf K.\ Heilmann}
\author[c]{Stephen Bongiorno}
\author[d]{Eric M.\ Gullikson}
\author[e]{Andrea Santangelo}
\author[e]{Chris Tenzer}
\author[f,g]{Ruth Kelly}
\author[f]{Silvia Zane}
\author[g]{Roberto Taverna}
\author[g]{Roberto Turolla}
\author[h]{Michela Negro}
\author[j]{Rozenn Boissay-Malaquin}
\author[k]{Luigi Gallo}
\author[e]{Honghui Liu}
\author[a]{Swati Ravi}
\affil[a]{MIT Kavli Institute for Astrophysics and Space Research, Massachusetts Institute of Technology, Cambridge, MA 02139, USA}
\affil[b]{Space Nanotechnology Laboratory, MIT Kavli Institute for Astrophysics and Space Research, Massachusetts Institute of Technology, Cambridge, MA 02139, USA}
\affil[c]{NASA Marshall Space Flight Center, Huntsville, AL 35805, USA}
\affil[d]{Lawrence Berkeley National Laboratory, Berkeley, CA 94720, USA}
\affil[e]{Eberhard Karls University of T\"{u}bingen, T\"ubingen, Germany}
\affil[f]{University College London, London, UK}
\affil[g]{Universita degli Studi di Padova, Padua, Italy}
\affil[h]{Louisiana State University and A\&M College, Baton Rouge, LA, 70803, USA}
\affil[j]{University of Maryland Baltimore County, Baltimore, MD 21250, USA}
\affil[k]{St.\ Mary's University, Halifax, NS, Canada}

\authorinfo{Further author information: (Send correspondence to H.L.M.)\\H.L.M.: E-mail: hermanm at mit.edu, Telephone: 1 617 253-8573}

\begin{document}
\maketitle

\begin{abstract}
 
The Globe Orbiting Soft X-ray Polarimeter (GOSoX) is a spectropolarimeter for the soft X-ray band.  GOSoX is based on the Rocket Experiment Demonstration of a Soft X-ray Polarimeter (REDSoX\footnote{Used with permission of the Red Sox Baseball Club and Major League Baseball.}), a NASA-funded sounding rocket payload. Like REDSoX, GOSoX consists of Wolter I X-ray optics from NASA/MSFC, critical-angle transmission (CAT) gratings made at MIT, and multilayer (ML) coated mirrors from LBNL that polarize the X-rays.  Colleagues at U. T\"ubingen will adapt commercially available sCMOS sensors for the focal plane.  The grating dispersion is matched to the lateral grading of ML mirrors set at 45\deg.  GOSoX can measure polarization across the entire 0.2-0.4 keV band with a spectral resolution E/dE $\sim$100.  Minimum detectable polarizations (MDPs) of 3-10\% are expected for over a dozen targets in a one-year mission. The mission was selected by NASA for launch in 2030.
 
\end{abstract}


\section{Introduction}

\label{sec:intro}

The Imaging X-ray Polarimetry Explorer (\ixpe, \cite{ixpe}) has demonstrated that 
X-ray polarimetry is now a valuable probe of physics in the most extreme environments: the magnetic fields of pulsars \cite{2022NatAs...6.1433D,2022ApJ...940...70M} and magnetars \cite{ixpe_4u0142,ixpe_1708}, the relativistic jets of active galactic nuclei \cite{Liodakis2022,DiGesu2022} and gamma-ray bursts \cite{ixpe_grb211009a}, and the accretion disks around black holes in galactic nuclei \cite{2022MNRAS.516.5907M,2023MNRAS.519.6138M} and in X-ray binaries \cite{ixpe_cygx1}. Despite the great success of IXPE, its sensitivity is limited to 2--8 keV, completely excluding  the soft X-ray band below 1 keV. The additional spectral features typical of the soft X-ray band cannot yet be examined polarimetrically, giving us an incomplete view of these extreme environments. 

The Astro2020 Decadal report explicitly says that X-ray polarimetry is a key capability needed to understand relativistic jets.
Astro2020 white papers presented the science case for 0.2--50 keV polarimetry \citep{2019arXiv190409313K} and an instrument \citep{xpp} with a low energy channel based on \rs\ \cite{redsoxjatis,redsox2024,redsox2025}.
See the companion paper for the current status of \rs\ that has been in development for 2.5 years \cite{REDSoXSPIE2026}.
\sxp\ is a orbital SmallSat based predominantly on \rs, selected by NASA in the 2025 call for Astrophysics Pioneer proposals.
Here, we describe the \sxp\ mission, providing
differences between \rs\ and \sxp\ instruments in section~\ref{sec:payloaddesign}, and the expected performance. 

\section{Science Objectives}


\subsection{Polarization Terms and Metrics}

\label{sec:metrics}

We refer to the polarization fraction as $\Pi$, the ratio of the linearly polarized flux to the total.  In Stokes formalism, $\Pi = \sqrt{Q^2 + U^2}/I$.
As commonly defined, the Minimum Detectable Polarization (MDP) is the lowest level of polarization that one can distinguish from random noise at the 99\% confidence level; when $\Pi >$ MDP, we would declare a detection of polarized flux.
For an instrument with modulation factor $\mu$,
MDP $= 4.29/(\mu R T^{1/2}) (R+B)^{1/2}$ \cite[cf.][]{weisskopf:77320E},
where $\mu$ is the ratio
of the polarization-modulated signal to the average for a 100\% polarized source,
$R$ is the source count rate, $T$ is the exposure time, and $B$ is the background rate.
When $B$ is small, as it is for our design (\S~\ref{sec:background}), then
MDP $= 4.29/(\mu [R T]^{1/2})$.  For a source photon flux, $n(\lambda)$, then $R = \int n(\lambda) A_{\lambda} d\lambda$, where $A_{\lambda}$ is the effective area of the instrument.  For a flat spectrum, $R = n_0 \int A_{\lambda} d\lambda \equiv n_0 {\mathcal A}$.  In this case,
MDP $\propto 1/(\mu {\mathcal A}^{1/2} )$, where ${\mathcal A}$ is the integrated area of the system (in cm$^2$\AA).

The MDP is an important metric for planning observations when one has no prior knowledge of the polarization angle, also known as the electric vector position angle (EVPA) as measured on the sky relative to celestial north.
Sometimes EVPAs can be set by external data.  Then, $\Pi$ can be estimated more directly, as was
done effectively in POGO$+$ observations of Cyg X-1 to
constrain the lamppost model of the accretion disk corona \cite{2018NatAs...2..652C} and in IXPE observations of Cyg X-1 \cite{ixpe_cygx1}.
In such a case, the uncertainty in $\Pi$ is
$\sigma_\Pi = 1 / [\mu (0.5 R T )^{1/2}] \approx $ MDP$/3$, so $\sigma_\Pi <3$\% for most of
our proposed observations.

In the next sections, we highlight some scientific investigations that are enabled by polarimetry in the 0.2-0.4 keV band of \sxp.

\subsection{Magnetized Atmospheres of Neutron Stars (NSs) and Pulsars}

The strongest magnetic fields in the universe are observed in isolated NSs -- up to $10^{14}$ G in the case of magnetars.
Thus, they are unique laboratories for the interaction between matter and magnetic fields, as well as for
testing quantum electrodynamics (QED) effects \cite{2006RPPh...69.2631H}.
With \sxp, we will sample a large range
of NS $B$-fields, observing  different targets with the goals of 
understanding the origin of spectral absorption lines, verifying QED predictions,
testing surface emission models, and assessing effects in pulsar magnetospheres.

The thermal surface emission from  many bright and close-by  isolated NSs  peaks below $1$ keV, which is outside \ixpe\ energy range but ideal for \sxp. 
Recent simulations \cite{redsoxKelly} have shown that \sxp\ can easily discriminate the composition of the surface layer and its phase state for such sources
(see \cite{redsoxKelly} and
\cite{2016MNRAS.459.3585G}).
Finding $\Pi > 30\%$ would clearly rule out the condensed surface scenario, so we target MDPs $<$ 30\% in each of 10 pulse phase bins, equivalent to a pulse averaged MDP $\le$ 10\%.
With NS spin periods typically in the $1$--$10$ s range \cite{2019A&A...622A..61S}, we also
require event time-tagging to $<$ 0.1 s.

Magnetars are also excellent laboratories for probing the presence of vacuum birefringence in regions of strong magnetic fields, as predicted by QED \cite{2002PhRvD..66b3002H,hsl03}. A smoking gun of this effect would be observing a very high degree of polarization from a large area of the NS surface.
RX J1856.6$-$3754 is a good target in this respect.  Given its very low pulsed fraction in the X-ray band \cite[$\approx 1.3\%$, see][]{2012A&A...541A..66S,2017MNRAS.465..492M}, we are observing emission from the entire surface at a constant temperature. Polarization measured at optical wavelengths in RX J1856.5$-$ 374 was suggested to be the first observational evidence of this effect \cite{2017MNRAS.465..492M}.
However, the VLT results cannot provide a definitive confirmation of the effect. The optical polarization was measured only at a confidence level $\sim 3\sigma$ and needs to be confirmed; in addition, plasma effects (e.g. dichroism) can be important at optical energies. 
As we expect the thermal emission in the \sxp\ band to come from a large fraction of the NS surface for isolated, cool NSs such as RX J1856.5-3754, the planned observations with \sxp\ can provide much stronger evidence of vacuum birefringence \cite{redsoxKelly}.

Several of the NSs that are bright only below 1 keV have
absorption features in their phase-averaged  spectra between $0.1$ and $0.5$ keV (see Tables~\ref{tab:targetsA} and \ref{tab:targetsB}). These features
are thought to arise from gravitationally redshifted O {\sc viii} \cite{2012MNRAS.419.1525H} transitions
or from proton cyclotron scattering/absorption
when $B \gae 10^{13}$ G \cite[e.g.][]{rxj0720,2007Ap&SS.308..181H,2017MNRAS.468.2975B}.
Spectro-polarimetric measurements will determine whether an absorption feature is atomic or cyclotronic.  If an absorption feature can be identified with an atomic transition, then the gravitational redshift at the NS surface, $z = (1-2GM/[Rc^2])^{-1/2} - 1$, can be computed providing the NS mass-to-radius ratio.
The polarization of a proton cyclotron line depends on the local magnetic field and varies with the pulse phase.
A cyclotron line is expected to appear as a dip in the polarization spectrum; simulations show that this is easily observable by \sxp, for magnetic field strengths $10^{13}$~G$ \lesssim B \lesssim  10^{14}$~G \cite{redsoxKelly}. 
Fig. \ref{fig:nsatm} shows a set of simulations of \sxp\ observations with different models, analyzed by the methods of \S~\ref{sec:datahandling}, showing that such a feature can show polarization different from the surrounding continuum.
While RX J1308.6$+$2127 also exhibits a deep absorption feature, its centroid varies with the pulse phase, likely produced in a small magnetic loop close to the surface \cite{2017MNRAS.468.2975B}.
Resolving these lines requires splitting the spectrum into $4$--$10$ bands,
with energy-integrated MDPs $<$ 10-15\% in order to obtain MDPs of $\le 30\%$  per band.

\begin{figure}
    \centering
    \includegraphics[width=\linewidth]{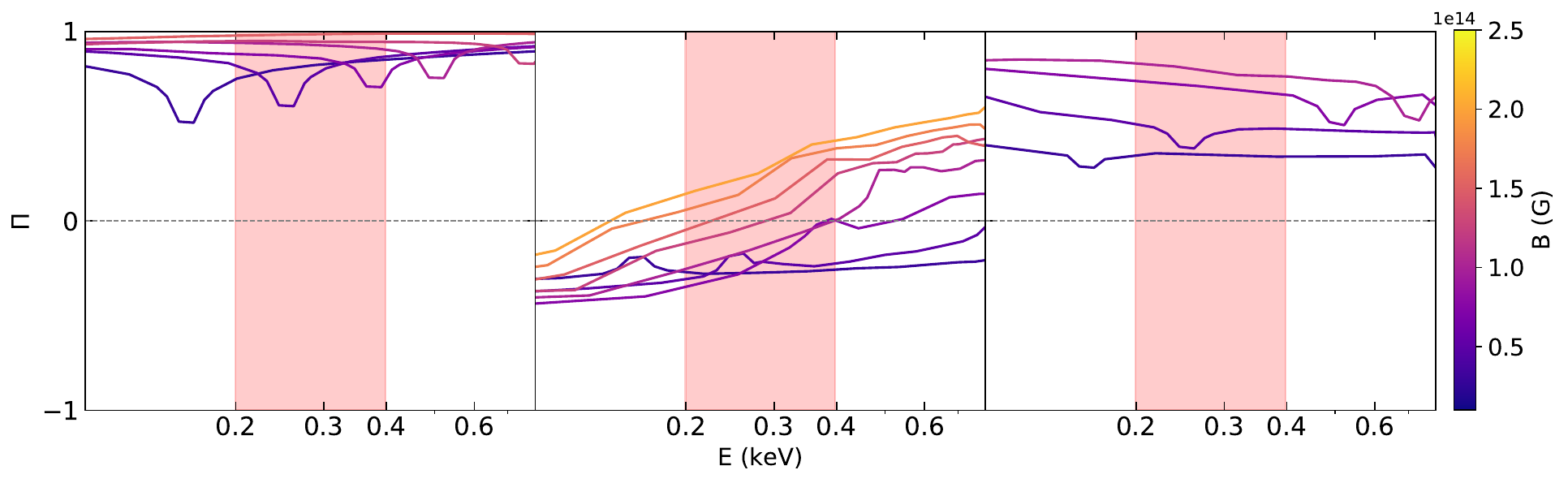}

    \caption{
    Results from three models of the polarization of an isolated neutron star's emission. Positive (negative) polarization fraction ($\Pi$) corresponds to X (O) mode dominated emission, with EVPAs at 90\deg\ to each other.  The shaded region corresponds to the \sxp\ band.  Magnetic fields are in units of $10^{14}$ G. {\em Left:} Pure plasma atmosphere, showing high polarization. Computation time limited the model spectral resolution, causing the absorption line to be shallower than expected.  {\em Center:} Plasma plus vacuum atmosphere assuming mode conversion. {\em Right:} Same as center but with partial mode conversion.  Note the various levels of polarization and that the proton cyclotron line is more weakly polarized than the continuum.  Figure adapted from \cite{redsoxKelly}.}
    \label{fig:rkfigs}
\end{figure}

\begin{SCfigure}
    \centering
    \includegraphics[width=8cm]{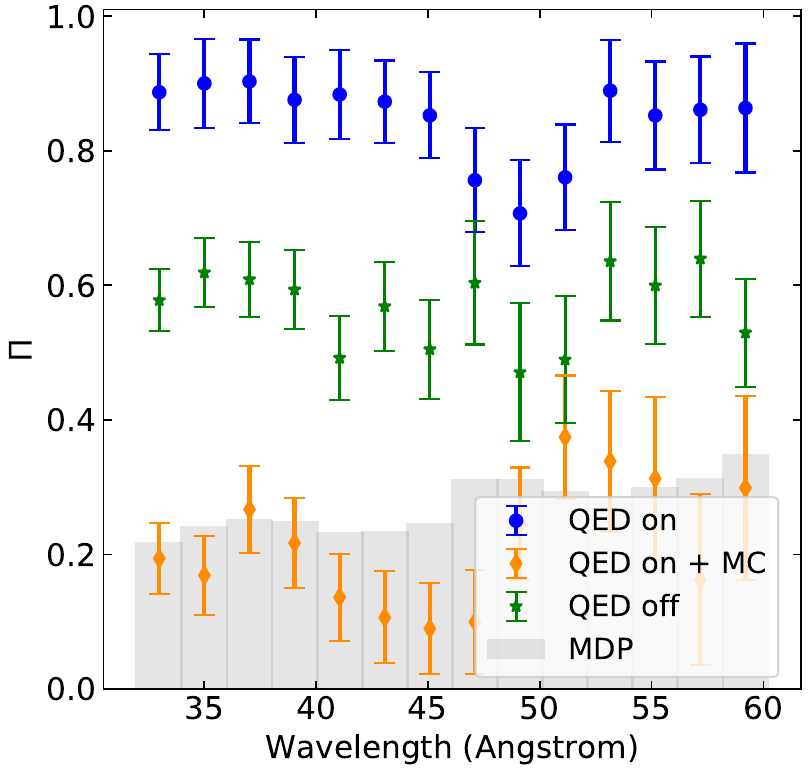}
\caption{
Simulations of a 0.5 Ms observation of RX J0720.4-3125, an isolated NS with an absorption feature
\cite{2012MNRAS.419.1525H}.  The blue points ($\pm 1 \sigma$) were simulated with a plasma plus vacuum atmosphere with no mode conversion at the vacuum resonance (Fig.~\ref{fig:rkfigs} left), the green points are a pure plasma model with QED turned off, and the orange points are for a plasma plus vacuum atmosphere with mode conversion (as in Fig.~\ref{fig:rkfigs}, center).  The absorption feature shows up in the QED$+$plasma model at better than 3$\sigma$.  Note that computation time limited the model spectral resolution, causing it to be shallower than expected. Figure adapted from \cite{redsoxKelly}.
}
\label{fig:nsatm}

\end{SCfigure}

Pulse-phased polarimetry provides the inclination of the pulsar spin axis, $i$, to the line of sight and the angle of the magnetic field to the spin axis, $\chi$, using the rotating vector model (RVM) \cite{2020A&A...641A.166P,ixpe_4u0142, 2023MNRAS.519.5902G}, originally applied to radio pulsars \cite{radha}.  Basically, the RVM gives the EVPA as a function of pulse phase, $\phi$, so fitting EVPA$(\phi)$ yields $i$ and $\chi$.
PSR B0656$+$14 and Her X-1 are ``traditional'' pulsars with $B \sim 10^{12}$ G.
An observation of PSR B0656$+$14 could distinguish between fan and pencil beam models of magnetospheric emission via rotation of EVPA with pulse phase using the RVM \cite{meszaros88}.
Pulsars have been observed by \ixpe\ but the X-ray emission of PSR B0656$+$14 peaks near 0.3 keV and is too faint above 2 keV for \ixpe.
The sub-keV pulsed emission of Her X-1 does not come from the surface or polar cap, as its effective size is much larger than the NS, and is more likely from the inner edge of the warped,
highly inclined accretion disk where the magnetosphere disrupts it \cite{2005ApJ...633.1064H}.
{Two scenarios can be distinguished using \sxp.
Either the beamed hard X-rays heat a spot on the disk, creating a scattering atmosphere polarized up to $\sim11\%$ \cite{chandra60} or the sweeping magnetic field heats the ionized disk and the radiation is predominantly synchrotron, polarized up to $\sim70\%$ if the field is very ordered there.}

\subsection{Soft X-rays from Active Galactic Nuclei (AGN)}

The soft X-ray emission of active galaxies can result from several
different mechanisms that can be distinguished with observations by \sxp,
ranging from strongly magnetized jets to mildly polarized accretion
disk atmospheres and coronae that current observations cannot discriminate.
The soft spectral components of these sources make \sxp\ an excellent
instrument to explore models of the soft X-ray emission.

\subsubsection{Relativistic Jets from Blazars}

\label{bllscience}

Blazars, which include BL Lac objects and highly variable quasars,
contain parsec-scale jets with
$\beta \equiv v/c \sim 0.98$ or higher.
Blazar jets are among the most powerful outflows in the universe.
In high-frequency synchrotron peak blazars (HSPs such as Mk 421),
the X-ray emission is most likely synchrotron radiation from high energy electrons. The 2-8 keV polarizations of both Mk 421 and Mk 501 are 10-15\%, 2-5 times stronger than the optical value \citep{Liodakis2022,DiGesu2022}. This agrees with models where the X-ray emission comes from a smaller volume upstream of the jet's more turbulent optical emission site. If so, X-ray polarization probes the more ordered magnetic field geometry where the jet flow is accelerated and collimated. Soft X-ray polarimetry with \sxp, alongside simultaneous optical and harder X-ray (IXPE) measurements, would fill the wavelength gap to provide a ``map'' of the magnetic field pattern and support the stratified shock model, if valid. 
Interpolating between the 15\% polarization in the 2-8 keV band and the 3\% value found in simultaneous R-band observations during the \ixpe\ observations \cite{DiGesu2022}, the estimated polarization in the \sxp\ band would be about 11\%, due to less ordering of the B field.
Thus, we target MDPs $\le$ 8\%.
In order to examine EVPA rotations possibly indicative of helical magnetic fields, such as found optically and with \ixpe\ for many sources \cite{2016A&A...590A..10K,2021MNRAS.507..225K,DiGesu2022}, HSP blazars will be reobserved over time scales of days to weeks.  For Mk 421 in particular, $\Pi = 0.1$ can be detected in as little as 4 ks, a much smaller time scale than achievable with \ixpe, testing the turbulent zone model \cite{2014ApJ...780...87M}.

Blazars with low-frequency synchrotron peaks (LSPs) like 3C 273 generally show weak X-ray polarization with limits ranging from 6-30\%, probably because their X-ray emission is dominated by Compton upscattering of unpolarized seed photons \cite{Ehlert22,2024ApJ...972...74M}.  3C 273, ``is neither fish nor fowl'' -- a very bright LSP that also has a soft excess like radio quiet AGN (\S \ref{agnscience}) and may be an example of an ``aborted jet'' \cite{2004A&A...413..535G}.

\subsubsection{Unabsorbed Seyfert 1 galaxies}

\label{agnscience}

Below $\sim 1$~keV, most, if not all, Type I AGN possess an additional component to the primary coronal power law.  The origin of this so-called soft excess is undetermined and vehemently debated.  
One possibility is that the soft excess originates from the blurred reflection component that also produces the relativistically broadened Fe~K$\alpha$ emission line at $6-7$~keV.  In this case, the blend of blurred, low-ionization lines in the reflection spectrum below $\sim 1$~keV creates the smooth excess we see \cite{2004MNRAS.353.1071F, 2005MNRAS.358..211R, 2006MNRAS.365.1067C}.
Another possibility is that the soft excess is a second Comptonizing region that is cooler and more optically thick than the conventional hot corona (e.g. \cite{1998MNRAS.301..179M, 2018A&A...611A..59P}).  If blurred reflection is present, then $\Pi$ can be $>$10\% \cite{2022MNRAS.510.4723P} and would be evident in the soft excess.  
The primary (hot) corona should contribute to the low-energy emission with $\Pi \sim 3-6\%$ between $0.1-1$~keV depending on the corona geometry (i.e. spherical or wedge) \cite{2017ApJ...850...14B}.  By contrast, a warm corona can be over 10\% polarized for a slab corona \cite{2022MNRAS.510.3674U}.  Thus, we target MDPs of 5\% to distinguish these models.

\subsubsection{Targets of Opportunity}

\label{sec:ToOscience}

\noindent
When a star approaches a supermassive black hole, a tidal disruption event (TDE) occurs if the tidal force becomes larger than the self-gravity of the star and if the radius when this happens is outside the black hole event horizon.  Part of the disrupted debris falls into the black hole from an accretion disk, giving rise to a bright UV or X-ray flare, which lasts for several months to a year. In 2011, two intriguing TDEs were detected by Swift. 
These TDEs have been interpreted as relativistic jets launched from central black holes \cite{2011Sci...333..203B, 2011Natur.476..421B}.
As with HSP blazars, the polarization of a stratified shock would be strongly dependent on energy, necessitating observations at soft X-rays with \sxp\ and at hard X-rays with \ixpe\ to test the model. 

\arrayrulecolor{myred!80!blue} 
\begin{table}
\begin{center}
\caption{Baseline Observing Plan, Priority A$^{\rm a}$ \label{tab:targetsA} }
\begin{tabular}{lrcll}
\rowcolor{myred!80!blue}
{\color{white}{\bf Category}} & {\color{white} {\bf Name}} &
   {\color{white} {\bf Time}}  &
   {\color{white} {\bf MDP}}  & {\color{white}{\bf Note}}\\
NSs & RX J1856.5-3754$^{\rm T}$	&		0.5	&	5	&		\\
 & RX J0720.4-3125$^{\rm T}$	&		0.5	&	6	&	293 eV abs'n line \cite{2012MNRAS.419.1525H}	\\
 & RX J1308.6+2127	&		0.4	&	7	&	107-256 eV line \cite{2017MNRAS.468.2975B}	\\
 & PSR B0656	&		0.2	&	7	&		\\
  & Her X-1$^{\rm T}$	&		0.3	&	3$^{\rm b}$	&	XRB	\\
\hline
Blazars & Mk 421$^{\rm T}$	&	0.3	&	2$^{\rm b}$	&	HSP	\\
 & Mk 501$^{\rm T}$	&		0.3	&	6$^{\rm b}$	&	HSP	\\
 & PKS 2155-304	&		0.3	&	6$^{\rm b}$	&	HSP	\\
 & TOO blazar	&		0.3	&	4	&	Flaring ISP	\\
\hline
AGN& Mk 478$^{\rm T}$	&		1.1	&	3	&	NLS1	\\
 & 1H 0419-577	&		0.8	&	3	&	Sy1.5	\\
 & TOO TDE	&		0.3	&	7	&		\\
\hline
Other & QQ Vul$^{\rm T}$	&		0.3	&	3	&	Polar	 \\
 & EX Hya	&		0.5	&	5	& Intermed. Polar	\\
 & Hz 43$^{\rm T}$	&	0.2	&	1.5	&	Null, WD	\\
\end{tabular}
\end{center}
$^{\rm a}${
    Exposure time is in Ms and MDPs are percentages. MDP is achievable with the listed exposure time and proposed instrument and MDP$_R$ is the required MDP.}
$^{\rm b}${
   MDP for each of three 100 ks observations.}
$^{\rm T}${
   Sources in the 3 month threshold mission.}
\end{table}

In intermediate synchrotron peak (ISP) blazars, inverse Compton (iC) emission begins to dominate the spectrum in the X-ray band.
The high energy tail of an LSP's synchrotron component extends into the soft X-ray band, so \sxp\ observations could show strong polarization even if the 2-8 keV spectrum dominated by the iC component is only weakly polarized \cite[cf.][]{2022ApJ...931...59P}.
ISPs often flare dramatically, necessitating a TOO to observe them in order to test the synchrotron nature of the soft X-ray flux with \sxp.

\subsection{Observatory Science}

\label{sec:otherscience}

The ROSAT Bright Source Catalog \cite[BSC,][]{rosatbsc} was examined for other potential \sxp\ targets.  Based on detailed modeling of a few sources and confirmed by computing area ratios, the \sxp\ count rate can be estimated as 0.0077 $R_s$, where $R_s$ is the ROSAT soft count rate.  The ROSAT ``soft band'' is centered at about 0.25 keV, which is in the middle of the \sxp\ band, so the correspondence is excellent.  The BSC provides the total count rate, $R$ and the hardness ratio $h$ that can be used to compute $R_s = R(1-h)/2$; sources with $h<0$, such as the blazars and the magnetars, are dominated by the soft band and are excellent targets for \sxp, depending on variability.  In the BSC, the sources in our priority A observing plan show up as the brightest of their respective classes.  There are $>200$ sources with $R_s > 1$ cnt/s in the BSC, for which \sxp\ can obtain an MDP of $<$10\% in 300 ks (using $\bar{\mu} = 0.936$, \S \ref{sec:lgmls}).

One type of source that shows up prominently in the BSC are accreting white dwarfs (WDs), generally known as cataclysmic variables (CVs).
A recent IXPE observation of EX Hya surprisingly showed that it was polarized at about 8\% in the 2-3 keV band but much less so above 3 keV \cite{exhya}.  This source is an intermediate polar, a type of CV with a magnetic field strong enough to funnel material from an accretion disk onto its poles.  The ultimate cause of the polarization is not certain, but it is likely to result from Thomson scattering of photons from the accretion column off the WD surface or from the outer layers of the column itself \cite{2004A&A...423..495M}.  The polarization depends on the geometry of the accretion column; thus \sxp\ opens a new avenue for testing such models.  EX Hya and several more strongly magnetic WDs in binaries (the so-called AM Her systems, also called polars) are bright in the soft X-ray band, as found in the ROSAT BSC.  We include the polar that is brightest in the soft X-ray band, QQ Vul.  Based on scattering models of other sources, we set the required MDP to 8\% but target values of 5\% or better.

\subsection{Observing Plan}

\label{sec:obsplan}

The baseline observing plan for a 6 month mission is given in Table~\ref{tab:targetsA} for an instrument with ${\mathcal A} = 35$ cm$^2$\AA, achievable with our design (\S~\ref{sec:prediction}).
Priority ``A'' sources comprise the prime mission, which will take 6 months at 40\% observing efficiency for a total of 6.3 Ms of available exposure time.
In addition to the science targets is a ``null'' calibrator, an isolated, non-accreting WD (HZ 43), which is expected to be unpolarized because its spectrum is dominated by thermal emission.  Other WDs are also bright in soft X-rays but HZ 43 is by far the brightest in the ROSAT BSC and has been used for calibration of the \axaf\ low energy transmission grating spectrometer \cite{2003SPIE.4851..157P,2006A&A...458..541B}.
The threshold mission consists of 1-2 sources from each category in the baseline plan and would take nominally three months to accomplish.
Priority ``B'' sources (Table~\ref{tab:targetsB}) are additional targets that could be observed under several circumstances, such as 1- having sufficient funding and schedule slack to operate longer than 6 months, 2- schedule gaps when A targets are not visible or have varied below reasonable brightness, or 3- if exposure times for A targets are reduced due to better instrument performance than expected.

The \sxp\ science team is still recruiting new members due to the variety of science targets. It currently consists of about 20 individuals who advise the PI regarding the mission science requirements and the mission observing plan.
The science team will be responsible for arranging joint observations with ground- and space-based telescopes to enhance the scientific return of \sxp{}.  Such observations are very important when observing variable sources.

\arrayrulecolor{myred!80!blue} 
\begin{table}
\begin{center}
\caption{Strawman Observing Plan, Priority B$^{\rm a}$ \label{tab:targetsB} }
\begin{tabular}{lrcll}
\rowcolor{myred!80!blue}
{\color{white}{\bf Category}} & {\color{white} {\bf Name}} &
   {\color{white} {\bf Time}} &
   {\color{white} {\bf MDP}}  & {\color{white}{\bf Note}}\\
NSs & RX J1605.3+3249	&		0.4	&	7	&	403 eV line \cite{2012MNRAS.419.1525H}	\\
  & Her X-1	&		0.3	&	3$^{\rm b}$	&	XRB	\\
\hline
Blazars & Mk 421	&	0.3	&	2$^{\rm b}$	&	HSP	\\
 & 1H 1426+427	&		0.3	&	6	&	HSP	\\
 & PKS 2155-304	&		0.3	&	6$^{\rm b}$	&	HSP	\\
 & TOO blazar	&		0.3	&	4	&	Flaring ISP	\\
 & 3C 273	&		1.2	&	3	&	LSP	\\
\hline
AGN  & Ark 564	&		1.1	&	4	&	NLS1	\\
 & Ton S 180	&		1.4	&	3	&	NLS1	\\
 & TOO TDE	&		0.3	&	7	&		\\
\hline
Other & VV Pup	&	0.2	&	5	&	Polar	\\
 & Hz 43	&		0.2	&	1.5	&	Null, WD	\\
\end{tabular}
\end{center}
$^{\rm a}${
   See notes to table~\ref{tab:targetsA}.}
$^{\rm b}${
   MDP for each of three 100 ks observations.}
\end{table}

\section{Payload Design}

\label{sec:payloaddesign}

\begin{figure}
    \centering
    \includegraphics[width=16.3cm]{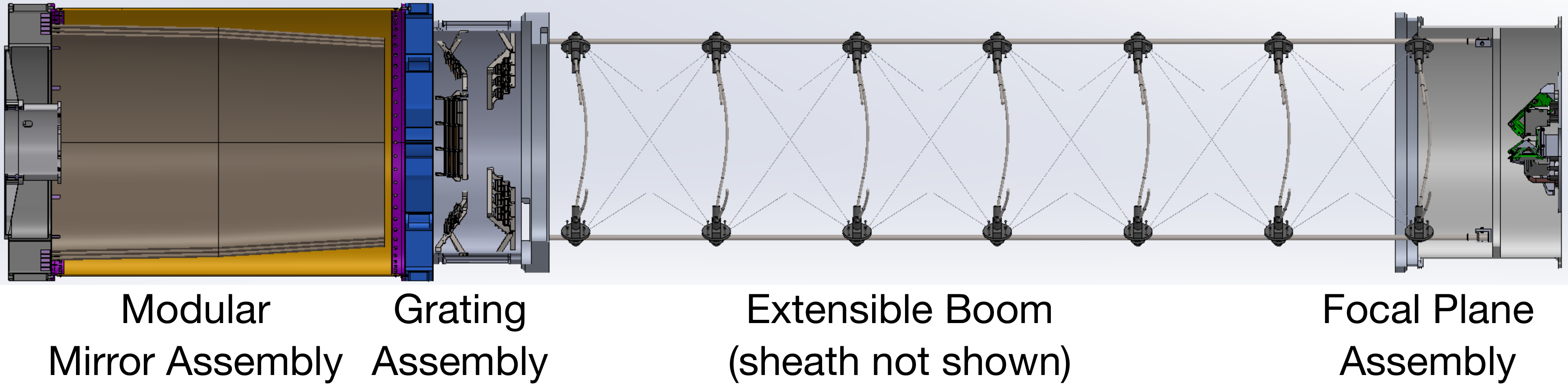}
 \caption{ 
Cutaway rendering of the deployed configuration for the \sxp\ instrument.
The star tracker is mounted inside the MMA volume at the forward end of the optics.
More details of the optics section and the focal plane
are shown in Fig.~\ref{fig:redsox_aperture}. The payload $+z$ axis is along
the optical axis, with the origin at the imaging detector, and $+x$ is
toward and along one of the LGMLs.  The boom design is under review.}
\label{fig:shortsxp}

\end{figure}

A rendering of the instrument is shown in Fig.~\ref{fig:shortsxp}.
As with \rs, the instrument consists of a dispersive X-ray spectrometer matched to
laterally graded multilayer (LGML) mirrors that polarize X-rays over a broad, soft X-ray band.
The LGML coatings provide high reflectivity at $45 \pm 5$\deg\ and polarization sensitivity.
The redirected, polarized X-rays are measured by sCMOS-based sensors.
The intensities in each of the three detector channels gives the Stokes parameters $I$, $Q$, and $U$.
The detectors also measure X-ray event positions along the dispersed, polarized
spectra, enabling polarimetric spectroscopy.

There are two main differences between the \rs\ and \sxp\ designs: 1- the \sxp\ optical bench has a boom that extends by 1.6 m after launch, and 2- the \sxp\ detectors are based on sCMOS sensors.

\subsection{Focusing Optics}

\label{sec:optics}

\label{sec:opticsdesign}

The optics group at the Marshall Space Flight Center (MSFC) will fabricate the mirror shells for the \sxp\ mirror module assembly (MMA). Three spare mirror shells from the \rs\ project will be used for the \sxp.
These are nested, confocal, Wolter I shells that are $l = 600$ mm long with an outer diameter of $r_{o} = 425$ mm.  The focal length of the MMA is $F =$ 2500 mm, for a focal plane plate scale of 82\arcsec{}/mm. At  graze angles of 1-1.2\deg, reflectivities average 88\%. We then reduce the geometric area by 12\% to account for shadowing by the assembly spider, giving a total effective area of 162 cm$^2$ that is approximately independent of energy in the 0.2-0.4 keV band.

\begin{figure}
   \begin{center}
   \begin{tabular}{c}
   \includegraphics[width=8.cm]{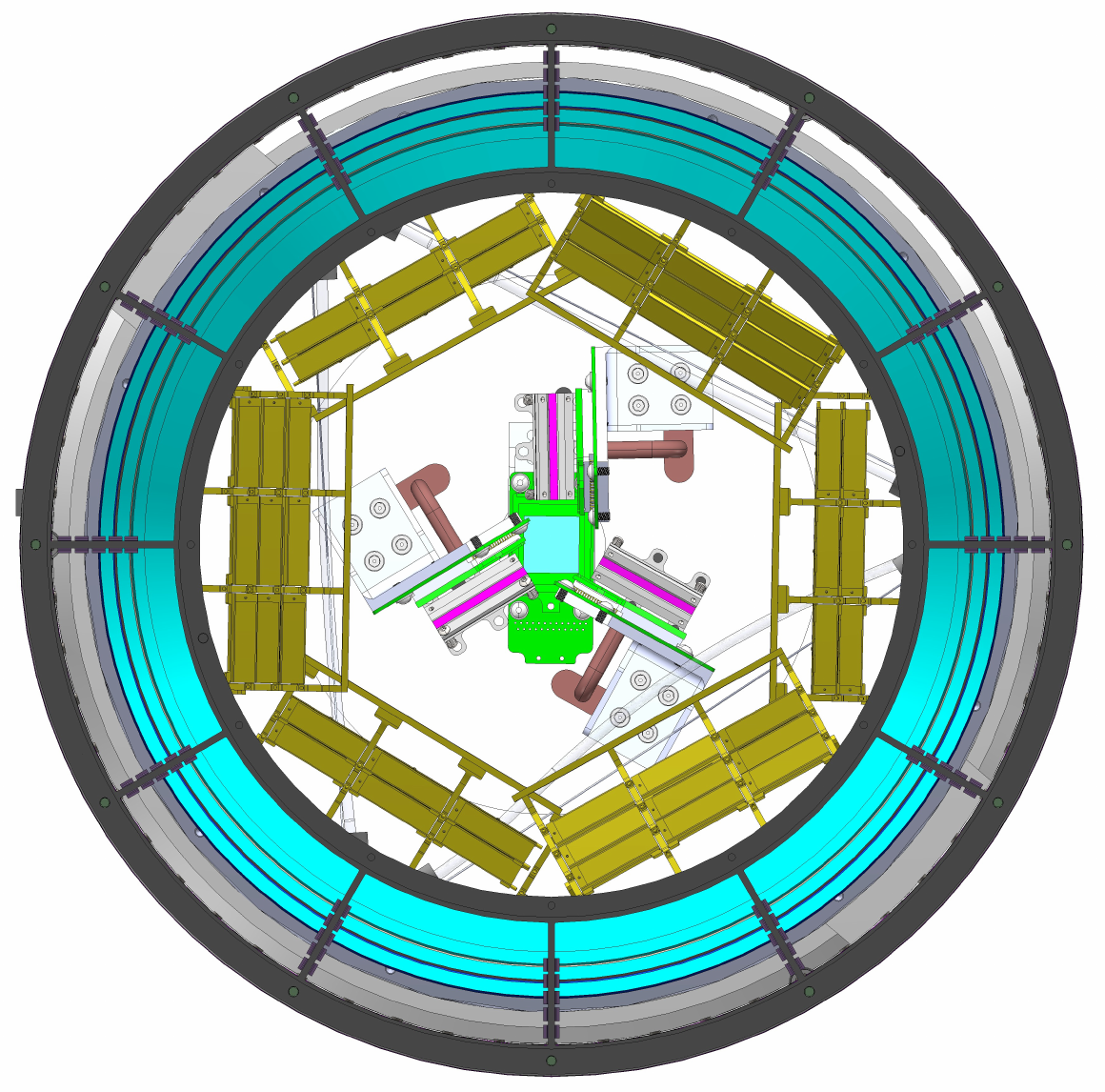}
    \includegraphics[width=8.cm]{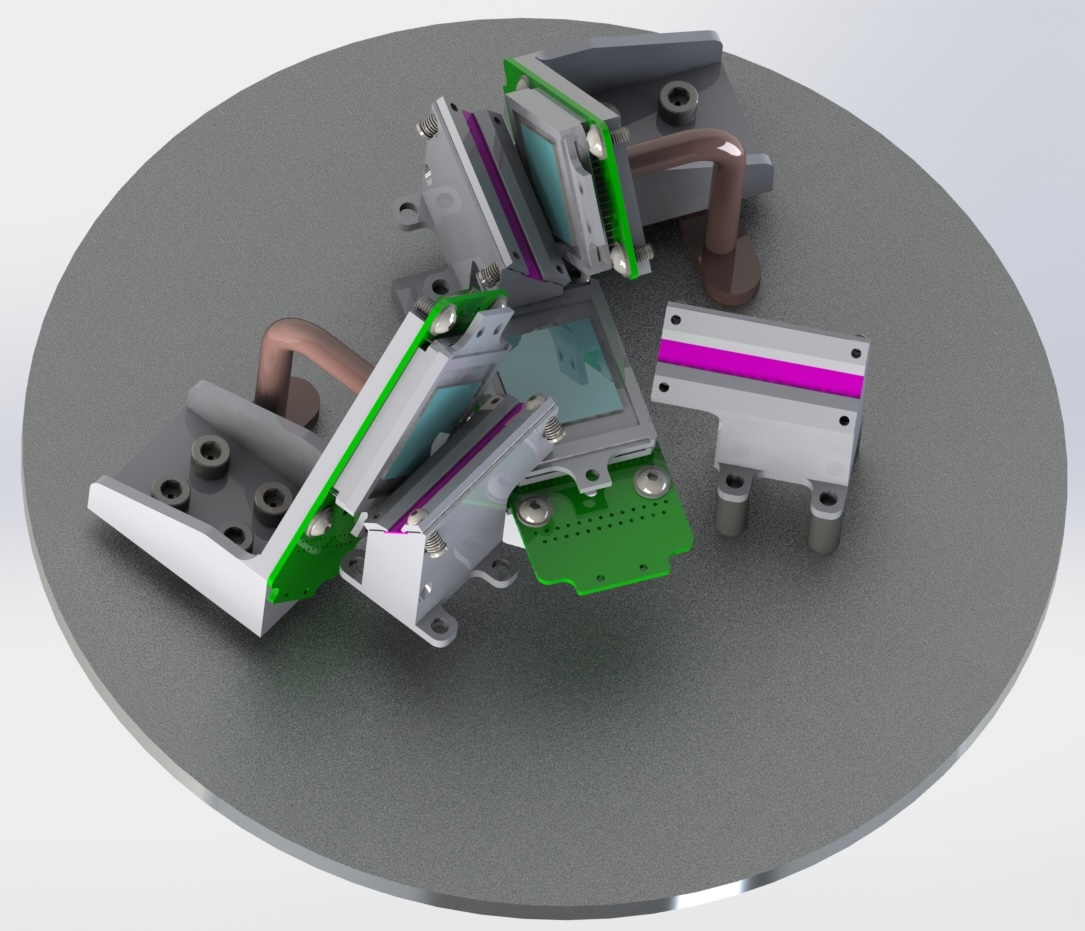}
 \end{tabular}
   \end{center}
 \caption{ 
{\it Left:} View of \sxp\ front aperture, without the central obscuring disk so that the gratings and focal plane are visible.
    The imaging detector (light blue) is along the optical axis of the mirrors (cyan) in the MMA; the polarimetry detectors are edge-on and not readily visible in this view.
    The gratings (yellow) are stair-stepped with three rows of gratings in the high ``petals'' (closer to the optics) and two rows in the low petals. (See \S\ref{sec:catgratings} \cite{redsoxjatis}.)  The grating sets in the left and right sectors disperse to the top LGML (magenta).
{\it Right:} Focal plane layout or \sxp, less one detector, similar to that of \rs.
  The imaging detector is slightly behind the MMA's imaging focus while the centerlines of the LGMLs are in the spectroscopy focal plane and reflect to the vertically oriented polarimetry detectors.  Copper straps connect the detector TECs to the spacecraft's cooling system.}
\label{fig:redsox_aperture}
\end{figure}

The spectral resolution of a transmission grating spectrometer is determined by the size of a point source's image along the dispersion -- only the 1D FWHM of the image is critical to the performance.
We used the Interactive RayTrace (IRT) code for IDL
to make spot diagrams, compute half-power diameters (HPDs), investigate off-axis aberrations, etc.
We define a spectral ``channel'' with opposing pairs of $\pm 30\deg$ sectors of mirror shells (Fig.~\ref{fig:redsox_aperture}).
This approach reduces the telescope 2D PSF for each channel from 25\arcsec\ HPD to a Gaussian profile with $\sigma_{\rm 1D} = 4.4$\arcsec, corresponding to $\sigma_{\rm PSF} = 54$ $\mu$m at the detector along the dispersion \cite[cf.][]{1987ApOpt..26.2915C}.
The front aperture is divided into three such pairs of sectors at 120\deg\ to each other.

The MSFC team will fabricate the MMA, assembling and aligning shells that were electroformed on highly polished mandrels. The procedure
is nearly identical to that used for FOXSI \cite{2016JAI.....540005C}, ART-XC \cite{2018SPIE10699E..1YP}, and IXPE \cite{ixpe}.

\subsection{Gratings}

\label{sec:catgratings}

The dispersion of a grating spectrometer is given by the grating equation: $\lambda = P \sin \alpha$, where $P$ is the grating period, $\alpha \approx x/F$ is the dispersion angle, and $x$ is the dispersion distance in the focal plane for a specific instrument channel.
We use the small angle approximation, as $x/F \ll 1$.
In order to cover the 0.2-0.4 keV bandpass (about 30-60 \AA)
onto a single detector of size $\Delta x = 38$ mm (see \S\ref{sec:ccds}),
the grating period, $P$, must be $\sim F \Delta \lambda /\Delta x \simeq 200$ nm.
Furthermore, the gratings should not have a plastic backing that absorbs soft X-rays, as used on \axaf\ high energy gratings \cite{hetgs}.
Critical-Angle Transmission (CAT) gratings \cite{2009SPIE.7437E..14H}
satisfy both requirements, with $P = 200$ nm and an open support structure built into the grating.

CAT gratings have been reliably produced for \rs\ in a 10$\times$30 mm format \cite{2015SPIE.9603E..14H,heilmann17}, identical to that needed for \sxp.
Dozens of CAT gratings
have been tested with soft X-rays
\cite{REDSoXSPIE2026}.
Efficiencies of $\sim$15\% were achieved in lab measurements in first order as needed for \sxp\ \cite{REDSoXSPIE2026}.
Alignment is performed at the mount and assembly level, as described in \S~\ref{sec:attitude}
\cite{2017SPIE10399E..15S}.
This method has been shown to align gratings to a single reference within the needed tolerances
based on analysis and raytracing \cite{redsoxjatis}.

A central feature of the instrument
design is matching the dispersed wavelength to the wavelength of
peak Bragg reflectivity of a LGML (\S~\ref{sec:lgmls}).
The multilayer reflectivity peaks at wavelength $\lambda = 2 d \cos \theta$,
where $d$ is the multilayer period and $\theta$ is the angle of incidence relative
to the LGML normal.
For \sxp, $d = G x$ (lateral grading), where $G$ is the
(constant) gradient of the multilayer's period along the grating dispersion.
If $z$ is the axial distance from the grating from the focal plane, then
$\lambda = P \sin \alpha \approx P x/z = 2 d \cos \theta$, giving
$z = P/(2 G \cos \theta)$.
Mounting the LGML at 45\deg\ (the Brewster angle) to
the telescope's optical axis sets $\theta = 45 \pm (4.4-4.9)$\deg, depending
on the precise grating location. 
The maximum value of $\theta$, 49.9\deg, gives the maximum of $z$,
which must be less than $2200$ mm, a condition
that is satisfied for $G = 0.72 $\AA/mm, the LGML gradient we target for \sxp.
We then solve for $z$ for each grating, which depends most strongly on
the radial distance of that grating from the optical axis,
resulting in a stair-stepped grating mount, as illustrated in Fig.~\ref{fig:shortsxp}.
Precise positioning has been computed and validated using raytracing code \cite{redsoxjatis} for \rs, also used for \sxp.
We plan to have two rows of three gratings each in the three lower sectors and three rows of three each in the three upper sectors for a total of 45 gratings.

The CAT gratings are used in a ``blaze'' configuration, tilted so that $+1$ order is significantly brighter than $-1$ order.  The optimal blaze angle, 0.7\deg, was determined using a grating efficiency model folded through the instrument response using a sample source, Mk 421.
Including blockage by the gratings' internal L1 and L2 support structures, the average measured efficiency of \rs\ gratings that are identical to those that would be used for \sxp\ is 15.4\% \cite{REDSoXSPIE2026}.
Accounting for 17\% blockage by the grating assembly mounts, the average transmission into $+1$ order is 12.8\% across the \sxp\ band at the optimal blaze angle.
The grating mounts and ``petals'' (see Fig.~\ref{fig:redsox_aperture}) will be practically identical to those designed for \rs.

\subsection{Laterally Graded Multilayer Mirrors (LGMLs)}

\label{sec:lgmls}

LGMLs of Cr and Sc have been made for \rs\ that are suitable for our design \cite{redsoxlgmls,REDSoXSPIE2026}.
Because the gratings must be mounted closer to the optics than in the \rs\ design, \sxp\ will require a slightly smaller multilayer period gradient.
Any deviation from linearity of the actual Bragg peak location, $\delta \lambda$, will degrade the system throughput by a factor of $\exp(-[f/\xi]^2/2)$, where $f = \Delta \lambda / \lambda$ is the fractional deviation from linearity.  The \rs\ flight LGMLs have $f < 0.01$ \cite{REDSoXSPIE2026}; averaging over the bandpass reduces $f$ by about $\times$2, giving average losses due to nonlinearity of less than 20\%, so the goal is $f < 0.005$, for $<$10\% loss.  The linearity and reflectivity of each LGML will be calibrated at the ALS.

\subsection{Detectors}

\label{sec:ccds}

As in \rs, four detectors are planned.  Three detectors measure the dispersed, polarization-sensitive spectra reflected by the three LGMLs and comprise the three polarimetry channels.  The fourth will view the zeroth-order image.
The three polarimetry channels are arranged to provide optimally independent information for deriving Stokes  $I$, $Q$, and $U$ at any instant.
The imaging detector's flux measurement gives us a redundant measurement of Stokes $I$.

For \rs, CCD detectors are used.  For \sxp, sCMOS sensors are preferred.
To reduce background, energy resolution of order 100 eV is sufficient (see \S~\ref{sec:background}).
Due to the LGML, second order from the gratings contributes less than 10\% to the signal, so filtering orders is not critical for \sxp.
The polarimetry detectors are not placed at the system focus for mechanical reasons, broadening images to about 0.5-1.0 mm, spatial resolution of order 0.2 mm is all that is required.
The imaging detector is used to assure proper acquisition of the target by measuring the centroid of the target to better than $\sigma_{\rm 1D} = 22$ $\mu$m.
Count rates are low, so frame times can be $\sim 1$ s but for phase-resolved polarimetry of pulsars, X-ray events should be time-tagged to better than 100 ms.

The detectors and associated frame-processing electronics will be contributed by the Institute for Astronomy and Astrophysics T\"{u}bingen (IAAT) of the Eberhard Karls University of T\"{u}bingen (EKUT) in Germany, funded by the German Aerospace Center (DLR)  and the EKUT High Energy Astrophysics group led by Prof.\ A.\ Santangelo.
The sensors for the cameras are TBD; we have baselined the GSENSE400BSI sensor from Gpixel that is processed with the Gpixel ``PulSar'' technology. These sensors have measured quantum efficiencies (QEs) better than 90\% in the 200--500 eV band \cite{2020APExp..13a6502H}, making them highly desirable for soft X-ray detectors. The sCMOS sensor has 11 $\mu$m pixels in a 2048$\times$2048 format, so the active area is 22.5 mm on a side, sufficient to capture the dispersed spectra for the polarimetry channels diagonally. The sCMOS sensor is capable of a frame readout time of 21 ms, well below the 100-ms readout time required for phase-resolved polarimetry.
At these fast read rates, optical blocking filters are not necessary \cite{heine24_detectors}.

The detectors will be operated at -30C to minimize dark current.
This temperature will be controlled to $\pm$1C by a single stage thermo-electric cooler (TEC) on each sensor, removing heat to the spacecraft thermal system's hot side via thermal straps.
However, at -7C, the dark current is only 3.7 e\textsuperscript{-}/s/pixel, so we could operate at this higher temperature \cite{2020APExp..13a6502H,heine24_detectors}.
The measurement noise would then be $\sim 14$ eV (1 $\sigma$) in the 80 to 1000 eV range, satisfying the required energy resolution. Therefore, the required operating temperature is within the capability of a passive cooling system, which might only require thermal straps from the sensors to the spacecraft thermal system, to be studied further during Phase A.



\label{sec:electronics}

The total power draw is expected to be 26.6 W including a 25\% contingency.  During standby mode (when not observing) about 1 W per sensor is needed to maintain each sensor's SDRAM.
Occasionally, full frames will be needed to diagnose bias maps or an anomaly, requiring a downlink of 50 Mb per sensor.
Onboard electronics will control power to the sensors and electronics to avoid operation during periods of high background.


\subsection{System Considerations and Other Components}

This section describes other aspects of the design of \sxp.

\subsubsection{Structure}

\begin{figure}
    \includegraphics[width=16cm]{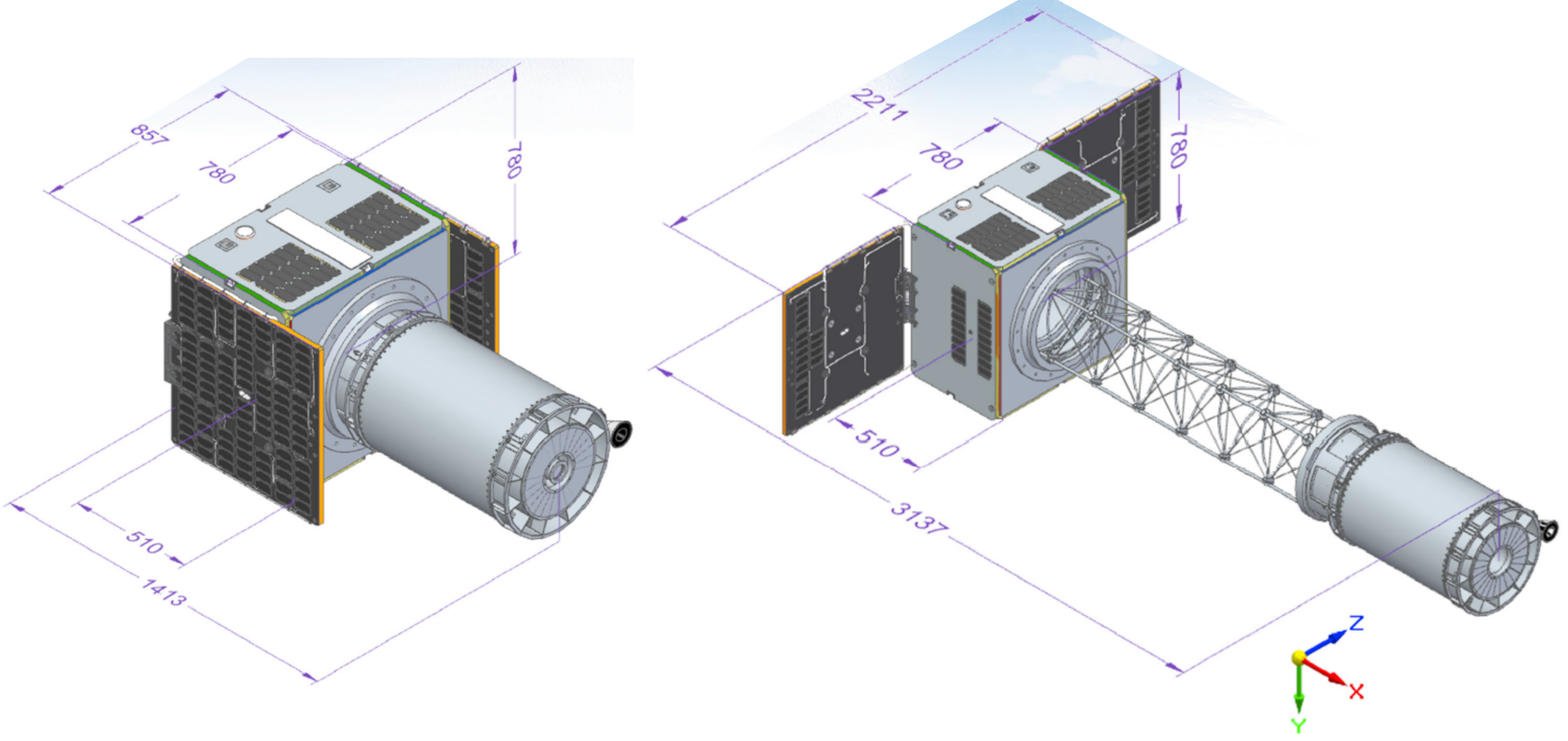}
 \caption{{
 Renderings of the SFL Dauntless spacecraft with \sxp\ payload in the stowed (left) and deployed (right) configurations. 
 The stowed dimensions of the satellite are are within the payload accommodation volume
(42\arcsec{} $\times$ 46\arcsec{} $\times$ 56\arcsec{})
, or $1067 \times 1168 \times 1422$ mm)
for a 24\arcsec{} ESPA ring diameter and a 5 m fairing (\S\ref{sec:launchandorbit}).  \sxp\ also conforms to the ESPA Grande payload requirements on mass ($<$ 465 kg) and center of gravity in the stowed configuration ($<$20\arcsec{} [500 mm] from the launch vehicle interface and $<1$ inch from the payload axis) (\S\ref{sec:launchandorbit}).  The sheathing and thermal layer around the payload are hidden.
  }}
\label{fig:spacecraft}
\end{figure}

The payload consists of two major sections. The fore section, the optics module, contains the mirror assembly and grating assembly.
The aft section (the focal plane assembly) contains the detectors, multilayer mirrors, and electronics systems.
A boom will
extend \sxp\ by 1.6 m from a 1.41 m length when stowed,
in order to fit into the ESPA Grande payload volume, to its operational length after launch (see \S~\ref{sec:scarchitecture}).
Figs.~\ref{fig:shortsxp} and \ref{fig:spacecraft} shows a notional boom; the current plan is to use a telescoping boom that is also light-tight, to prevent light leaks.

Tolerances were determined using analytical formulae
verified by raytracing \cite{redsoxjatis}.
Mechanical machining and assembly tolerances are sufficient
for most aspects of the system.
The alignment of the gratings to the grating assembly reference is carried out on an optical bench using a UV laser system
(\S~\ref{sec:gratingintegration}).

\subsubsection{Attitude Control and Alignment}

\label{sec:attitude}

Light would be lost due to attitude jitter if the dispersed
spectrum is displaced along the dispersion direction by an offset angle $\delta$,
because $\lambda$ will be displaced from the Bragg reflectivity peak.
Modeling attitude variations with a 1D Gaussian, the Bragg reflectivity loss is given by
$\eta = \sigma_{\rm LGML} (\delta^2 + \sigma_{\rm 1D}^2 + \sigma_{\rm LGML}^2)^{-1/2}$,
where $\sigma_{\rm LGML}$ is wavelength-dependent and depends on which gratings are
involved (high or low, \S~\ref{sec:lgmls}) and $\sigma_{\rm 1D} = 54$ $\mu$m (\S~\ref{sec:optics}) is the contribution of optics blur to the reflectivity loss.
Compared to $\delta = 0$, $\eta$ drops only 15\% at 31 \AA\ (the most sensitive part of the \sxp\ bandpass)
for $\delta = 90$ $\mu$m; at the center of the band, the loss is 8\%, indicative of the average loss.
Thus, as long as $\delta < 90$ $\mu$m, attitude jitter will be a minor contributor to effective area loss.
To provide a 1D blur of $<$ 90 $\mu$m, the attitude jitter should
then be $<$ 11\arcsec\ at 1$\sigma$, defined as the radius of the circle
that encloses 68\% of the telescope axis orientations.

In order to vary the position angles of the LGML mirrors
with respect to the sky as a hedge against channel-to-channel calibration errors,
the spacecraft roll angle will be varied to span a range of 360\deg\ by continuously rotating or stepping at periodic intervals through the observation.
In order to accurately assign the roll angle to any given X-ray event, the scan rate should be no faster than 1\deg/s and the roll angle should be known {\it post facto} to 1\deg\ every second during data collection.
The initial roll angle for any given observation is arbitrary.
See \S~\ref{sec:scarchitecture} for more details about the attitude system.

 \begin{SCfigure}
    \includegraphics[width=8cm]{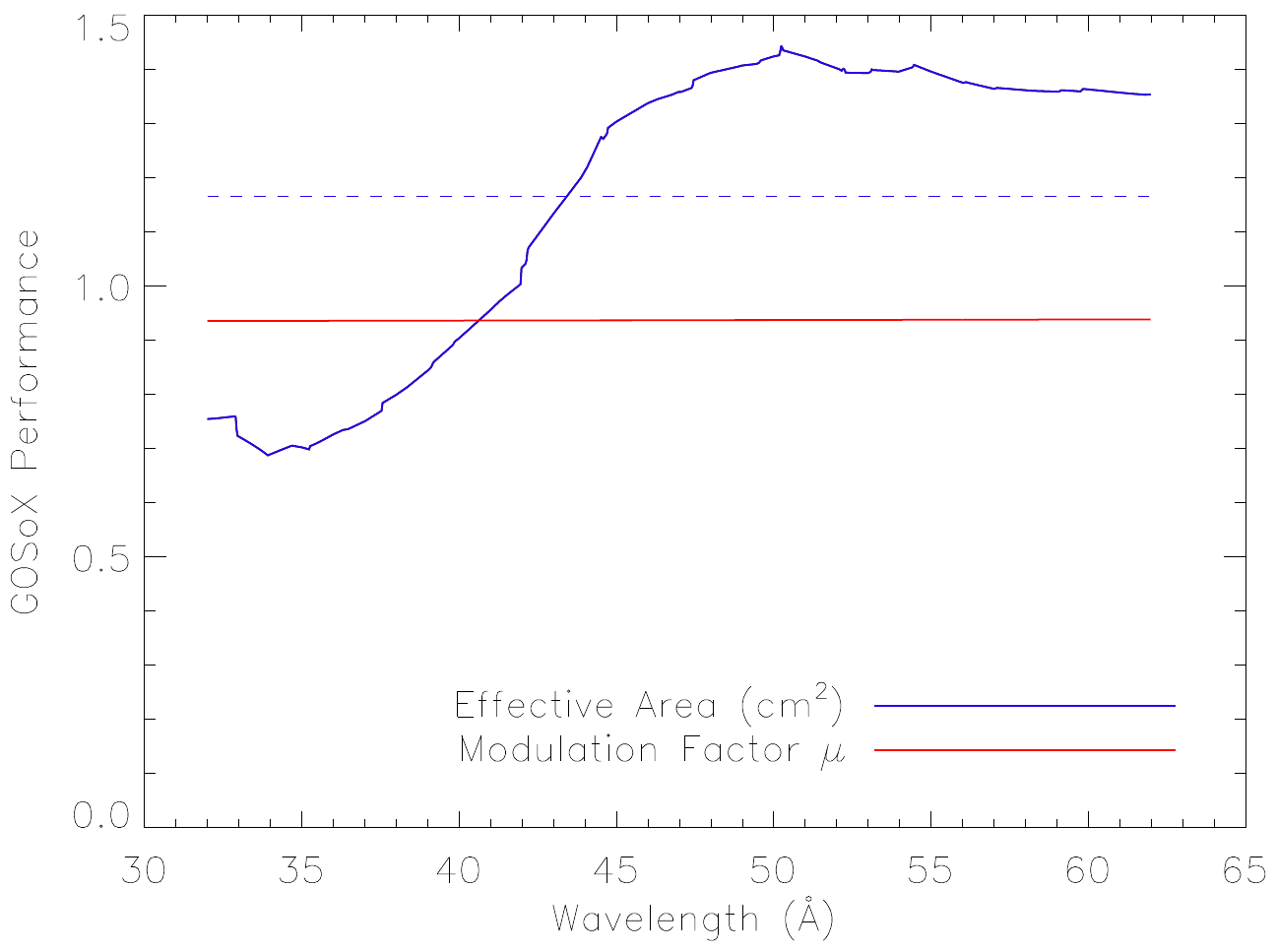}
 \caption{
    Expected \sxp\ system performance.
    {\it Blue:} Effective area of
  the system based on component performance  The average effective area is 1.165 cm$^2$ (dashed blue line).
  {\it Red:} Modulation factor, averaged between the values obtained for 40\deg\ and 50\deg\ graze angles.  The averaged modulation factor is 93.6\%, almost independent of wavelength.
  }
\label{fig:ea_mu}
\end{SCfigure}

\subsubsection{Sources of Background}

\label{sec:background}

From the Suzaku mission (600 km altitude and 31\deg\ inclination),
the particle background in the
backside-illuminated CCD was
$5 \times 10^{-8}$ cnt/s/keV/pixel, or
$1.5 \times 10^{-5}$ cnt/s/mm$^2$ in a 0.2 keV band at 0.3 keV \cite{2008SPIE.7011E..2CL}.
In the cross dispersion direction, the dispersed spectral extraction region will be two rectangles of size 0.4 $\times$ 22 mm, separated by about 2 mm.
The spectra's heights result from the narrow extent of the
mirror radii (about 20\% of their diameters).
The total extraction region is about 50 mm$^2$ for 3 independent
channels and 2 spectral zones, giving an expected
particle background of 0.0008 count/s.
For an exposure of $10^6$ s, there should be $<$800 counts from particles;
obtaining an MDP of 10\% requires about 2,500 counts, so particle background
is minor but included in our MDP estimates in any case.
The X-ray background in the \sxp\ bandpass is dominated
by Galactic emission. A full ray-trace simulation including in particular gratings and the multi-layer mirror using the {\tt marxs} code \citep{2017AJ....154..243G,redsoxjatis} show that off-axis X-rays are dispersed in such a way that their reflection is suppressed by the ML, in contrast to on-axis photons which are diffracted such that they hit the Bragg-peak. Using a measured flux for the Galactic emission and background AGN \citep{2009PASJ...61.1117B}, the ray-trace predicts an X-ray background rate of 2 counts per 1 Ms -- clearly negligible. 

\subsubsection{Data Handling}

\label{sec:datahandling}

The count rate in each detector will be measured after selecting for the
expected energy range based on the grating dispersion
and using detector resolution to reduce background.
The expected source count rate in angle bin $i$, with
azimuthal angle $\psi_i$ (relative to celestial North) is
$R_{i} = \int [ I +  \mu (Q \cos 2\psi_i + U \sin 2\psi_i) ] A(\lambda) d \lambda$, where $I$, $Q$, and $U$ have units of ph/cm$^2$/s/\AA.
The three Stokes parameters are used to determine the fraction of linearly polarized
light, $\Pi = (Q^2 + U^2)^{1/2} / I$, and the polarization angle $\phi \equiv (1/2)\arctan U/Q$, both as a function of energy.
For a flat spectrum source, we find

\begin{equation}
\begin{aligned}
I& = (R_1 + R_2 + R_3) / {(3{\mathcal A})}\\
Q& = (2 R_1 - R_2 - R_3) / {(3  \mu {\mathcal A})}\\
U& = (R_2 - R_3) / {(3^{1/2}  \mu {\mathcal A})}
\end{aligned}
\label{eq:stokes}
\end{equation}

\noindent
using the observed count rates $R_i$ in detector $i$ (as a function of $\lambda$), and referencing to
where detector 1 happens to be
aligned to celestial east \cite{redsoxjatis}.
Given sufficient signal, the Stokes parameters can be measured as a function of wavelength.
Usually, analysis using instrument line response functions (LRFs) would require a forward-folding methodology \cite[cf.][]{2017ApJ...838...72S} but the LRFs for a spectrometer with energy-resolving detectors are nearly diagonal, making direct analysis more straightforward.

Eqs.~\ref{eq:stokes} appear to require that $\mu {\mathcal A}$ be independent of channel, which would require careful cross-calibration of each instrument channel in both polarized and unpolarized light.
However, our simulations with variations up to $\pm$10\% between channels show that by rotating the instrument $\pm 30$\deg, we can actually cross-calibrate values of $\mu {\mathcal A}$ for the channels with a full likelihood analysis of any in-flight data as long as $\Pi$ and $\phi$ does not change systematically during spacecraft rotation. Estimates of $\Pi$ and $\phi$ and their uncertainties are essentially unaffected.  Furthermore, $\mu$ is identical for each channel due to the design of \sxp, so we only need to verify relative effective areas at a large beamline (see \S~\ref{sec:IandT}) or with the in flight observations of null calibration sources.

\subsubsection{Baseline Performance}

\label{sec:prediction}

\begin{SCfigure}
     \includegraphics[width=8cm]{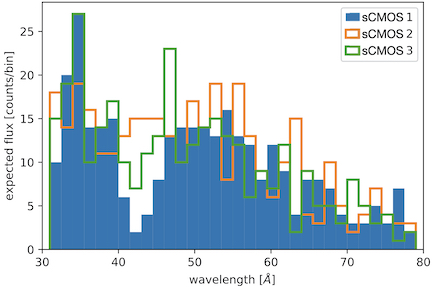}
 \caption{{
 Simulation of a \sxp\ observation of a hypothetical NS, showing each channel's count spectra.  The continuum is unpolarized except in an absorption feature at 43\AA\ (290 eV) that is 5 \AA\ (30 eV) wide with an equivalent width of 2.5 \AA\ (15 eV) and 50\% polarized along a position angle that nulls the flux in detector 1.  There are only 1000 counts in the observation, showing that a polarized absorption feature can be detected using a spectropolarimeter of our design.  For this simulation, the bandpass of the instrument was extended to 80 \AA.
}}
\label{fig:simdata}
 \vspace{-0.2in}
\end{SCfigure}

We used detailed models of the grating efficiency ($\epsilon_\lambda$), LGML reflectivity ($r_\lambda$), QE curves,
and raytracing to compute the integrated area: ${\mathcal A} = \int A_\lambda d\lambda = A \int  \epsilon_\lambda \eta_\lambda r_\lambda Q_\lambda d\lambda$. For the nominal attitude jitter, $\eta_\lambda \approx 1-e^{-(\lambda+\tilde{\lambda})^2/\Lambda^2}$, where $\tilde{\lambda} = 37$ \AA\ (27 \AA) and $\Lambda^2 = 2800$ \AA$^2$ (2000 \AA$^2$), for the low (high) grating spectra.  We obtain ${\mathcal A} = 35$ cm$^2$\AA.
One may also estimate ${\mathcal A}$ by $\bar{A} \Delta \lambda$, where $\bar{A} = 1.165$ cm$^2$ (Fig.~\ref{fig:ea_mu})
and $\Delta \lambda = 30 $\AA\ is the bandwidth of the system, giving ${\mathcal A} = 35$ cm$^2$\AA.
The zeroth order count rates for our targets are 0.2-20 cnt/s.
{\tt Marxs} raytraces \citep{2017AJ....154..243G,redsoxjatis} of a source with a polarized absorption feature demonstrates the value of a polarimeter with a broad spectral band with spectral resolution such as provided by \sxp\ (Fig.~\ref{fig:simdata}).

\subsection{Integration and Testing}
\vspace{-0.2cm}
\label{sec:IandT}
The instrument integration and test plan consists of three main parts: component testing, subsystem testing and integrated instrument tests.  Component-level testing with X-rays of the detectors, gratings, the MMA, and the LGMLs will be the responsibilities of the various institutions fabricating them.
Payload- and satellite-level vibration testing, thermal vacuum testing, and Electromagnetic compatibility (EMC) testing will be performed at MSFC, followed by closeout testing and burn-in. The spacecraft will then remain in launch-ready storage at MSFC before delivery for integration with the launch vehicle.

System integration begins when the detectors and LGMLs are mounted on the focal plane.  Independently, the gratings are installed on mounts and integrated and aligned in sectors that comprise the grating subassembly (\S~\ref{sec:gratingintegration}).
This unit will then be integrated with the MMA in two phases.
In the first phase, the optics will be offset forward of the grating and focal plane assembly using spacers to account for the finite distance of the unpolarized X-ray source at the MSFC 100 m beamline.
In this configuration, the uniformity of the effective area with channel and the alignment of the MMA can be verified.  A polarized source at the C-K$\alpha$ line at 277 eV will also be used to verify polarimetric performance.
Then, the spacers are removed for final assembly with the boom in the stowed configuration and environmental testing, which includes vibration tests and thermal vacuum cycling and balance.

\subsubsection{Component Testing: Gratings}
\label{sec:gratingintegration}
Grating quality is verified in the MIT Polarimetry beamline as for \rs\ \cite{REDSoXSPIE2026}.
Upon mounting, we will use a grating alignment procedure that was developed for and demonstrated as part of Arcus phase A work \cite{2017SPIE10399E..15S,heilmann17} and set up for \rs.
This technique uses a UV laser and position-sensitive detectors to measure the reflection and back diffraction off of the CAT grating bars.  The orientation and alignment of the gratings can be determined and adjusted within the tolerances required by \sxp\ as an assembly, all on an optical bench.

\subsubsection{Focal Plane and Grating Assembly Integration}
\label{sec:fplaneintegration}

Integration and alignment of the focal plane assembly, which includes all four detectors and LGMLs, requires a determination of the LGML position as mounted on the focal plane with respect to the imaging detector.  
Synchrotron measurements provide accurate Bragg peak positions with respect to an LGML edge that will be used as a reference for mechanical assembly and metrology using a coordinate measuring machine (CMM). 
The positioning can be verified using the MIT Polarimetry beamline.
The grating sub-assembly (described in \ref{sec:gratingintegration}) will be aligned to the focal plane assembly using the alignment cubes of the two subsystems on an optical bench.

\subsubsection{Optics Testing}
\label{sec:opticsintegration}
The MMA will be tested at the MSFC 100 m beamline to validate the effective area model and verify the optics HPD.
For these measurements, the optic will be mounted on a PI H-850-H2V hexapod and illuminated with 0.28 keV line emission.
The HPD will be measured with a soft X-ray sensitive CCD (Andor DW436) and the effective area will be measured with a pair of Amptek Fast-SDDs with Si$_3$N$_4$ windows.
In order to examine the 1D image profiles and measure the azimuthal uniformity of the effective area, the entrance aperture will be blocked with sector masks and the PSF measured using the facility CCD.

\subsubsection{Integrated Instrument Testing}
\label{sec:envirotesting}


We plan to test in the MSFC 100 m beamline where the source is about 100 m from the focal plane.
By offsetting the mirror with spacers about 64 mm forward from the flight position, the instrument
focal length will match that of the flight instrument for the purposes of testing with the finite source distance.
While the reflection angles off the MMA surfaces will change,
the gratings will retain their flight distances from the LGMLs, ensuring that the Bragg
condition is satisfied as for flight.
Thus, alignment and relative effective area can be tested in the beamline.
The spacer lengths can be precisely determined and then removed for final instrument-level environmental testing (including vibrational and thermal vacuum tests) at the MSFC environmental test facility.
 
 \label{sec:contamination}
 
 The optics and detectors are sensitive to contamination.  An accumulation of 10 nm of
 nonvolatile residue (NVR) would reduce mirror reflectivity by 20\%, increasing
 MDPs by 10\% (e.g., from 8.0\% to 8.8\%).  For the detectors, it would take
 100 nm of NVR to cause a similar decrease in sensitivity.
 While a nonvarying neutron star could
 be used to monitor contamination variation, our approach is to prevent
 such buildup.


\section{Spacecraft}
\label{sec:spacecraft}
\subsection{Architecture}

\label{sec:scarchitecture}

The Space Flight Laboratory (SFL) at the University of Toronto Institute for Aerospace Studies will supply the spacecraft for \sxp.
SFL has excellent experience with NASA Pioneer programs, providing two of the spacecraft for approved (and continuing) Pioneer missions, Aspera \cite{2021SPIE11819E..03C} and StarBurst. As a result, they are familiar not only with the development of spacecraft for such missions, but also US export control requirements.
The \sxp\ spacecraft is designed for low Earth orbit (LEO), based on their Dauntless bus.
The structure will be composed of a baseplate (below which the ESPA ring separation system is mounted), a thrust structure
and a top plate that stabilizes the instrument within the spacecraft.

\subsection{Communications}
\label{sec:communications}

The Telemetry \& Command (T\&C) subsystem consists of an SFL S-band command receiver (up to 32 kbps), a pair of SFL S-band patch antennas mounted on opposite faces of the spacecraft to provide omni‐directional coverage, and an SFL S‐band Telemetry Transmitter (500mW RF output power) nominally operated at 2 Mbps.
At an instrument data rate average of 10 kbps and 40\% observing efficiency (\S~\ref{sec:ops}), we expect to generate at most 45 MB per day from the instrument.
With two contacts of 6 minutes per day to one ground station and a throughput of 80\% (accounting for packet overhead and transmission losses), the S-band system can handle up to 144 MB per day.

\subsection{Attitude Control}

In order to maintain the alignment of the dispersed spectrum and the Bragg
peak of the LGML, the pointing axis
knowledge and control requirements are
set at 11\arcsec{} at 1 $\sigma$, converting to
a 1D blur on the detector of $<$ 35 $\mu$m (see \S\ref{sec:attitude}).
There are two aspects of the attitude control budget
for the spacecraft attitude control system: short term
variations over seconds to hours and fixed offsets.
The former are not correctable in-flight so they contribute to the pointing jitter. 
The fixed offsets will be calibrated in flight during the commissioning phase of the mission that begins when the telescope door is opened.
Slews of 90\deg\ only require 1-2 minutes, so we plan to orient the solar panels
to the sun to recharge batteries during target occultation every orbit.
For more power-positive targets, maneuvers can take place within each orbit to optimize observing efficiency.  The star tracker has a 22\deg\ solar avoidance angle, so \sxp\ has an anti-sun field of regard of at least $2\pi$ sr.
The star tracker has 5.5\arcsec\ absolute performance at 1 $\sigma$. When combined with ADCS control overshoot and reaction wheel jitter, the system performance changes slightly to 5.6\arcsec\ for short term errors and another 2.4\arcsec\ for longer term errors from gravity gradient and other disturbance torques (based on flight experience), giving a combined 1 $\sigma$ jitter of 8.0\arcsec, thus meeting our requirements.



\section{Mission Implementation}

\subsection{Launch Vehicle and Orbit}

\label{sec:launchandorbit}

The baseline mission involves operations in LEO for one year, noting that the threshold mission can be accomplished in less time (\S~\ref{sec:obsplan}).
An injection altitude of 400-500 km will provide this orbital lifetime.
The best observing efficiency is for an equatorial orbit (like IXPE currently occupies), which can be as high as 60\%. In a 28\deg\ inclination orbit, particle background will also be higher as the instrument passes through the South Atlantic Anomaly, reducing observation time by about 30\%, giving a net observing efficiency of 42\%.
Thus we assume that the observing efficiency is at least 40\%.

\subsection{Operations}

\label{sec:ops}

Operations costs are based on two S-band contacts per day (\S~\ref{sec:communications}).
The Mission Operations Center (MOC)
is responsible for scheduling contacts, uploading
commands, and storing downlinked data onto a secure server.
Due to daily contacts, schedule changes can be accommodated with 2-3 days, in response to TOOs.
Data will then be available for download to the \sxp\ Science Operations Center (GSOC) at MIT.
The GSOC will be responsible for the long-term observing plan that is used to create the detailed
observing plan on a monthly basis.
The GSOC will develop the on-board quaternion schedule to be transferred securely to the MOC for uplink to the flight computer.
The GSOC will collect data from the MOC and convert housekeeping and
engineering data to physical units for assessing the health of the
instrument.

\subsection{Data Management}

\label{sec:dmp}

Data processing and analysis will be performed at the GSOC.
There are several basic data products starting
with L0, which is raw telemetry from the ground station
that includes housekeeping and spacecraft attitude data.
L1 data are event lists and timelines of quantities
that are used for data analysis such as telescope
roll angle, earth angle, position in orbit, etc.
L2 are event lists filtered to include only data
from good time intervals when the detector was exposed to the source
and acceptable grades, defined during in-orbit commissioning.
L3 data include source maps (from the imaging detector), spectra,
catalogs of source positions and fluxes, and light curves.
The catalogs will include such information as dates of observations,
polarization fractions (or limits) and EVPAs
for significant detections.
Source observations may last 20 days or more,
comprising about 1 GB of data in event lists.

The GSOC will also maintain an active archive, calibration database and observation catalog.
We will follow the approach taken by the \ixpe\ project
and will provide all data in FITS formats compatible with software
developed by the \ixpe\ project for fitting X-ray polarization data.
All L1, L2, and higher-level data, as well as the software and
calibration data required for producing these products from
the L1 data, will be archived at the HEASARC according to
established HEASARC archiving practices within two months of
the completion of an observation.

\section{Technology Development}

Several systems of the \sxp\ instrument would be flown in orbit for the first time on this mission
and are vital to future missions.  Thus, \sxp\ would raise the TRLs of these components.
CAT gratings with 200 nm periods were developed for Arcus \cite{heilmann17,2017SPIE10397E..0QS}, proposed as a Probe Class mission, and as an option for Lynx \cite{heilmann17}.
Arcus was designed to provide high-resolution soft X-ray spectroscopy in the 12-50\AA\ bandpass with sensitivity orders of magnitude higher than any previous astronomical observatory.  Among its science goals are to trace the propagation of outflowing gas from the disks of AGN.
Lynx is a large X-ray imaging telescope being considered as a flagship mission that would follow on from the success of \axaf\ \cite{lynx}.
Lynx is designed to ``enable highly challenging observations in each of three broad science pillars: detecting and understanding the seeds of the first supermassive black holes, characterizing physics of the energetic processes that drive galaxy formation and evolution, and probing the broad range of high-energy processes that shape stellar birth and death, internal stellar structure, star-planet interactions, and the origin of elements'' \cite{2019SPIE11118E..0JB}.

This project would put sCMOS detectors into orbit for use below 0.5 keV.  
Some X-ray cameras with sCMOS sensors (and the GSENS400BSI in particular) are already available commercially such as the Sydor Wraith, the Andor Marana X, and the pco.edge 4.2 bi XU.  CMOS sensors were flown on a NASA-funded sounding rocket for solar observations \cite{2018NIMPA.912..191I,2022A&A...665A.103B} and on the Einstein Probe \cite[0.5-4.0 keV,][]{EinsteinProbe}.  These detectors are all based on sensors from Gpixel.  Such detectors have great promise for use in orbital missions as an alternative to CCDs due to lower power requirements, faster frame rates, and higher operational temperatures.

Laterally graded multilayer mirrors (LGMLs, \S~\ref{sec:lgmls}) have never been flown and are the critical polarizing element of our design, which is the only broad-band polarimetry method in the sub-keV band of which we are aware.
Just as \ixpe\ is an Explorer to measure polarization in the 2-8 keV band,
\sxp\ is a Pioneer of the sub-keV band.
The X-ray Polarization Probe (XPP, \cite{xpp}) would
combine larger scale versions of these instruments
with a 10-50 keV polarimeter, such as demonstrated by
X-Calibur \cite{10.1117/1.JATIS.4.1.011004}.
Thus, XPP would be a general observatory, capable of measuring polarizations
across the 0.1-50 keV band simultaneously.
Current lab development funded by the NASA APRA program is focused on extending LGML usage up to 0.8-1.0 keV, thereby more thoroughly covering the sub-keV band.
The \sxp{} project will demonstrate the scientific value of such a mission and the LGMLs needed for the soft polarimetry channel.
See the XPP white paper \cite{xpp} and the associated
science white paper \cite{2019arXiv190409313K} for more details.

\section{Plans}

\label{sec:phasea}
We plan for five development phases over the five year project: development of the Project Plan (Phase A, 6 months), preliminary design (Phase B, 6 months after a 3 month delay), manufacturing of the engineering model and detailed design (Phase C, 10 months),
flight model manufacturing and its integration and testing through launch and commissioning (Phase D, 21 months), flight operations and project closeout (Phase E, 8 months), with built-in schedule margin of six months that can be used as needed.
Phases A and B are comparatively short due to the significant prototyping and design work in common with and available from the \rs\ project.

\acknowledgments 
The \sxp\ MIT team is supported by NASA grant 80NSSC26M0046 to MIT.

\bibliographystyle{spiebib} 
\bibliography{apj-jour,polarimetry26}

\end{document}